\documentclass{emulateapj}

\usepackage{graphicx}
\usepackage{float}
\usepackage{amsmath}
\usepackage{epsfig,floatflt}

\renewcommand{\d}[0]{\mathbf{d}}
\newcommand{\n}[0]{\mathbf{n}}
\newcommand{\s}[0]{\mathbf{s}}
\renewcommand{\a}[0]{\mathbf{a}}
\newcommand{\m}[0]{\mathbf{m}}

\newcommand{\F}[0]{\mathbf{F}}
\newcommand{\T}[0]{\mathbf{T}}
\newcommand{\C}[0]{\mathbf{C}}
\renewcommand{\L}[0]{\mathbf{L}}

\newcommand{\N}[0]{\mathbf{N}}
\newcommand{\M}[0]{\mathbf{M}}
\newcommand{\iN}[0]{\mathbf{N}^{-1}}
\newcommand{\iM}[0]{\mathbf{M}^{-1}}
\newcommand{\w}[0]{\mathbf{w}}
\renewcommand{\S}[0]{\mathbf{S}}
\renewcommand{\r}[0]{\mathbf{r}}
\renewcommand{\u}[0]{\mathbf{u}}
\newcommand{\q}[0]{\mathbf{q}}
\renewcommand{\v}[0]{\mathbf{v}}
\renewcommand{\P}[0]{\mathbf{P}}

\newcommand{\di}[0]{d_i}

\newcommand{\fknee}[0]{f_{\textrm{knee}}}

\begin{document}

\title{Bayesian noise estimation for non-ideal CMB experiments}

\author{I. K. Wehus\altaffilmark{1,2}, S. K. N{\ae}ss\altaffilmark{3}
  and H. K. Eriksen\altaffilmark{3,4}}

\email{i.k.wehus@fys.uio.no}
\email{sigurdkn@astro.uio.no}
\email{h.k.k.eriksen@astro.uio.no}

\altaffiltext{1}{Theoretical Physics, Imperial College London, London
  SW7 2AZ, UK}

\altaffiltext{2}{Department of Physics, University of
  Oslo, P.O.\ Box 1048 Blindern, N-0316 Oslo, Norway}

\altaffiltext{3}{Institute of Theoretical Astrophysics, University of
  Oslo, P.O.\ Box 1029 Blindern, N-0315 Oslo, Norway}

\altaffiltext{4}{Centre of Mathematics for Applications, University of
  Oslo, P.O.\ Box 1053 Blindern, N-0316 Oslo, Norway}


\begin{abstract}
We describe a Bayesian framework for estimating the time-domain noise
covariance of CMB observations, typically parametrized in terms of a
$1/f$ frequency profile. This framework is based on the Gibbs sampling
algorithm, which allows for exact marginalization over nuisance
parameters through conditional probability distributions. In this
paper we implement support for gaps in the data streams and
marginalization over fixed time-domain templates, and also outline how
to marginalize over confusion from CMB fluctuations, which may be
important for high signal-to-noise experiments. As a by-product of the
method, we obtain proper constrained realizations, which themselves
can be useful for map making. To validate the algorithm, we
demonstrate that the reconstructed noise parameters and corresponding
uncertainties are unbiased using simulated data. The CPU time required
to process a single data stream of 100\,000 samples with 1000 samples
removed by gaps is 3 seconds if only the maximum posterior parameters
are required, and 21 seconds if one also want to obtain the
corresponding uncertainties by Gibbs sampling.
\end{abstract}

\section{Introduction}
\label{sec:introduction}

Detailed observations of the cosmic microwave background (CMB) during
the last two decades have revolutionized cosmology. Through detailed
measurements of the angular CMB power spectrum, a highly successful
cosmological concordance model has been established, stating that the
universe is statistically isotropic and homogeneous, filled with
Gaussian random fluctuations drawn from a $\Lambda$CDM spectrum, and
consists of 4\% baryonic matter, 23\% dark matter and 73\% dark energy
\citep[e.g.,][ and references therein]{komatsu:2011}. Using this model,
millions of data points from many different types of cosmological
observations can be fitted with only six free parameters.

This success has been driven primarily by rapid progress in CMB
detector technology, allowing experimentalists to make more and more
detailed maps of the CMB fluctuations. However, such maps are
imperfect, in the sense that they typically are contaminated by
various instrumental effects. For instance, the optics of a given
experiment can be asymmetric; the detector gain may be unknown and
time-dependent; the data may exhibit resonant frequencies due to
electronics or cooling non-idealities; and the observations are
invariably noisy. All these non-idealities must be properly understood
before one can attempt to extract cosmology from the observations.

In this paper we consider one specific component within this global
calibration problem, namely how to estimate the statistical properties
of the instrumental noise in light of real-world complications. This
problem has of course already been addressed repeatedly in the
literature \citep[e.g.,][]{prunet:2001,hinshaw:2003}, and our method
is in principle similar to that of \citet{ferreira:2000}, taking a
Bayesian approach to the problem. The main difference is that we
formulate the algorithm explicitly in terms of a Gibbs sampler
including both the time stream and the noise parameters as unknown
variables, and this has several distinct advantages. First, it allows
us to obtain proper uncertainties on all derived quantities. Second,
gap filling is directly supported through built-in proper constrained
realizations. This can for instance be used to account for
instrumental glitches in the time stream, or to exclude point sources
and other bright sources from the analysis. Third, it is
straightforward to add support for additional nuisance parameters, due
to the conditional nature of the Gibbs sampler. In this paper we
implement template marginalization, which may for instance be useful
for removing cosmic ray glitches in the Planck HFI data
\citep{planck_hfi:2011} or ground pickup for ground based experiments
\citep[e.g.,][]{quiet:2011}. We also outline the formalism for
marginalization over CMB fluctuations, which may be relevant for
experiments with high signal-to-noise ratio.

The method presented here is mathematically identical to the CMB Gibbs
sampling framework developed by
\citet{jewell:2004,wandelt:2004,eriksen:2004,eriksen:2008}, and the
main difference is simply that the object under consideration is a
one-dimensional time stream instead of a two-dimensional field on the
sphere. This makes the implementation considerably simpler, and the
run times correspondingly faster. As a demonstration of the
practicality of the method, we apply it to simulated data with
properties typical for current ground-based experiments, and
demonstrate explicitly that the computational costs of the method are
tractable. The experiment of choice will be \citet{quiet:2011},
for which this method was initially developed.

\section{Data model}
\label{sec:model}

The first step of any Bayesian analysis is to write down an explicit
parametric model for the observations in question. In this paper, we
start with the assumption that the output, $\d$, from a given detector
can be written in terms of the following sum,
\begin{equation}
\d = \n + \P\s + \T\a+ \m.
\label{eq:model}
\end{equation}
Here each term indicates a vector of $n$ values sampled regularly in
time in steps of $\Delta t$, that is, $\d = \{\di\}$ with $i=1,\ldots,N$.

The first term on the right-hand side, $\n$, indicates the
instrumental noise, which is our primary target in this paper. All the
other components are only nuisance variables that we want to
marginalize over.

We assume that the noise is Gaussian distributed and stationary over
the full time range considered. In practice this means that the full
data set of a given experiment should be segmented into parts which
are individually piecewise stationary. For QUIET this corresponds to
division into so-called ``constant elevation scans''
\citep{quiet:2011}, while for Planck it corresponds to division into
so-called ``rings'', which are one-hour observation periods with a
fixed satellite spin axis \citep{planck:2011}. Because the noise is
assumed stationary, the time-domain noise covariance matrix, $\N$,
depends only on the time lag between two observations, $N_{tt'} =
N(t-t')$: It is a Toeplitz matrix, and may therefore be well
approximated in Fourier domain with a simple diagonal matrix, $N_{\nu\nu'}
= N_\nu \delta_{\nu\nu'}$. Here $N_\nu$ is the Fourier-domain noise power
spectrum, which is given by the Fourier transform of $N(t-t')$.

Our main task is to estimate $N_\nu$, and we do so in terms of a
parametrized function. For many experiments this function is well
approximated by a so-called $1/f$ profile,
\begin{align}\label{eq:Nnu}
N_\nu=\sigma_0^2\left[1+\left(\frac{\nu}{\fknee}\right)^\alpha\right]
\end{align}
which describes a sum of a correlated and an uncorrelated noise
component in terms of three free parameters. The white-noise RMS
level, $\sigma_0$, defines the overall amplitude of the noise; the
knee frequency, $\fknee$, indicates where the correlated and the
uncorrelated components are equally strong, and $\alpha$ is the
spectral index of the correlated component. Collectively, we denote
$\{\sigma_0, \alpha, \fknee\}$ by $\theta$. Of course, other
parametrizations may easily be implemented if necessary.

The second term on the right-hand side, $\P\s$, indicates the
contribution from the CMB sky, with $\P$ being a pointing matrix,
typically equal to zero everywhere except at $P_{ip}$ if the detector
points towards pixel $p$ at time $i$, and $s_p$ is the true (beam
convolved) CMB signal. We make the usual assumption that $\s$ is
isotropic and Gaussian distributed with a given angular power
spectrum, $C_{\ell}$. In this paper, we will simply outline the
formalism for how to deal with this term, and leave the implementation
for a future paper dedicated to Planck analysis; as mentioned in the
introduction, this machinery was initially developed QUIET, which is
strongly noise dominated for a single data segment, and the CMB
component is therefore not important, as will be explicitly
demonstrated in this paper.

The third term is a sum over $n_{\textrm{temp}}$ time-domain
templates. These can be used to model several different types of
nuisance components. Three examples are diffuse foregrounds and cosmic
ray glitches for Planck, and ground pick-up for QUIET. In either case,
we assume in this paper that the template itself is perfectly known,
and the only free parameter is an overall unknown multiplicative
amplitude $a$. This is a vector of length $n_{\textrm{temp}}$, and $\T$
is the two-dimensional $n \times n_{\textrm{temp}}$ matrix listing all
templates column-wise.

Finally, the fourth term on the right-hand side of Equation~\ref{eq:model} denotes a time-domain mask, $\m$. This is implemented
by a ``Gaussian'' component having zero variance for samples that are
not masked, and infinite variance for samples that are masked. In
order to make analytic calculations more transparent, we write the
the corresponding covariance matrix as a diagonal matrix with elements
\begin{equation}
M_{ii} = \left\{ 
\begin{array}{ll}
 a, & i \textrm{ not masked}  \\
 \epsilon, & i \textrm{ masked}  \\
 \end{array} 
\right.,
\end{equation}
where $a \rightarrow \infty$ and $\epsilon \rightarrow 0$. A typical
application of this component is to remove periods of instrumental
glitching, or to discard particularly bright observations when the
telescope points towards bright astrophysical sources, such as point
sources or the Galactic plane.

\section{Gibbs sampling and the posterior}
\label{sec:posterior}

Our primary goal is now to map out $P(\theta|\d)$, the noise spectrum
posterior distribution marginalized over all nuisance components. By
Bayes' theorem this distribution reads
\begin{equation}
P(\theta|\d) = \frac{P(\d|\theta)P(\theta)}{P(\d)} \propto
\mathcal{L}(\theta)P(\theta),
\label{eq:posterior}
\end{equation}
where $\mathcal{L}(\theta) = P(\d|\theta)$ is the likelihood,
$P(\theta)$ is a prior on $\theta$, and $P(\d)$ is an irrelevant
normalization constant. 

In this paper we adopt for simplicity uniform priors on $\sigma_0$,
$\alpha$ and $\fknee$. For typical relevant time series which contain
$\sim10^5$ samples, these parameters are usually strongly data-driven,
and the choice of priors is therefore irrelevant. However, if an
informative prior (or the Jeffreys' prior) is desired for a given
application, it is straightforward to include this as indicated by
Equation~\ref{eq:posterior}.

Since we assume that the noise is Gaussian distributed with covariance
$\N(\theta)$, the likelihood is given by
\begin{equation}
\mathcal{L}(\theta) \propto \frac{e^{-\frac{1}{2}\n^T
    \N^{-1}(\theta)\n}}{\sqrt{|\N(\theta)|}},
\label{eq:likelihood}
\end{equation}
where $\n = \d - \P\s - \F\a - \m$ is the noise component of the
data stream.The goal is to compute this
distribution, marginalized over $\s$ and $\a$, while at the same time
taking into account possible gaps in the data.

The latter point touches on an important computational issue. If there
are no gaps in the data, then $\N$ is a Toeplitz matrix, and
multiplication with $\N$ is performed most efficiently in Fourier
space. However, the same does not hold if there are gaps in $\d$,
since the symmetry of $\N$ is broken. The well-known solution to this
problem is to fill the gap with a constrained noise realization with
the appropriate spectrum \citep[e.g.,][]{hoffman:1991}. In our
formulation, this is equivalent to estimating $\n$ jointly with
$\theta$.

More generally, we want to estimate the joint density $P(\n, \m, \s, \a,
\theta|\d)$, from which any desired marginal may be obtained. At first
sight, this appears like a formidable computational problem, involving
more than $10^5$ free parameters. However, this is also a problem that
may be tackled by means of the statistical technique called Gibbs
sampling, which has already been described in detail for computing
the CMB angular power spectrum with contaminated data by
\citet{jewell:2004,wandelt:2004,eriksen:2004,eriksen:2008}. 

According to the theory of Gibbs sampling, samples from a joint
distribution may be obtained by iteratively sampling from each
corresponding conditional distribution. For our case, this leads to
the following sampling scheme,
\begin{align}
\label{eq:n}
\m,\n &\leftarrow P(\m,\n|\s, \a, \theta, \d) \\
\s,\n &\leftarrow P(\s,\n|\a, \theta, \m, \d) \\
\a,\n &\leftarrow P(\a,\n|\theta, \m, \s, \d) \\
\theta &\leftarrow P(\theta|\n, \m, \s, \a, \d) 
\label{eq:theta}
\end{align}
The symbol $\leftarrow$ indicates sampling from the distribution on
the right-hand side. With this algorithm, $(\n, \m, \s, \a, \theta)^i$
will be drawn from the correct joint distribution. 

Note that each of the sampling steps that involve time-domain vectors
are joint steps including the noise component itself. This approach is
highly computationally advantageous as it allows for fast
multiplication with $\N$ in Fourier domain; conditional algorithms for
sampling each component separately would require slow convolutions in
time domain. Of course, it is fully acceptable within the Gibbs
sampling machinery to sample some components more often that others.

Note also that if we are only interested in the joint
maximum-posterior parameters, we can replace the relevant steps in the
above algorithm by a maximization operation, such that we maximize the
conditional instead of sampling from it. The algorithm then reduces to
a typical iterative approach, but formulated in a convenient and
unified statistical language. The advantage of this approach is
computational speed, while the disadvantage is the loss of information
about uncertainties. Both versions of the algorithm will be
implemented and demonstrated in the following.

\section{Sampling algorithms}
\label{sec:algorithm}

Equations~\ref{eq:n}--\ref{eq:theta} defines the high-level
algorithm in terms of conditional sampling steps. To complete the
method, we have to establish efficient sampling algorithms for each
conditional distribution.

\subsection{Noise estimation with ideal data}
\label{sec:noise_estimation}

Perhaps the most fundamental conditional distribution in the sampling
scheme outlined above is $P(\theta|\n, \m, \s, \a, \d)$. This describes
the distribution of the noise parameters given perfect knowledge about
all components of the data. To obtain an explicit expression for this
distribution, we first note that $P(\theta|\n, \m, \s, \a, \d) =
P(\theta|\n)$; if we know the true noise component, $\n$, no further
information about either the CMB signal, the template amplitudes, or,
indeed, the actual data is needed in order to estimate the noise
parameters.

\begin{figure}[t]
\mbox{\epsfig{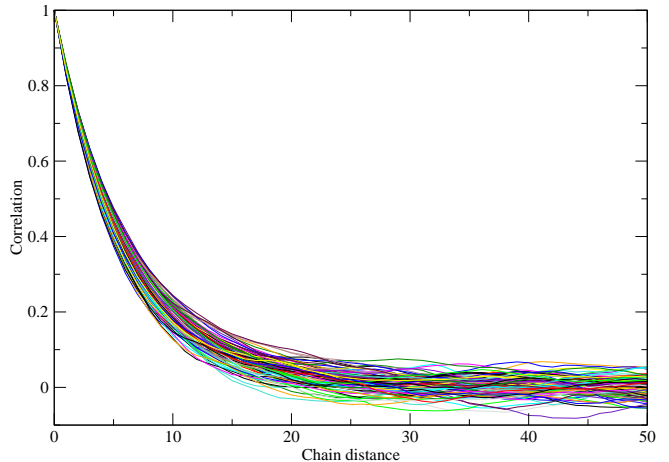}}
\caption{Correlation function for $\alpha$ of the Metropolis sampler employed to
  sample from $P(\theta|\n)$ with a curvature matrix based proposal
  density. Similar plots for $\sigma_0$ and $\fknee$ look visually the same. The correlations fall below 10\% at a lag of $\sim$20 samples,
  and we adopt a thinning factor of $\sim$20 samples to suppress
  correlations, given that the computational cost of this sampling
  step is lower than the constrained realization sampler.}
\label{fig:corrfunc}
\end{figure}

The expression for the conditional distribution $P(\theta|\n)$ is then
formally the same as that for $P(\theta|\d)$ given by
Equations~\ref{eq:posterior} and \ref{eq:likelihood}.  Writing this
out in Fourier space for the $1/f$ profile discussed in
Section~\ref{sec:model}, one finds the following explicit distribution
for $\theta=\{\sigma_0, \alpha, \fknee\}$,
\begin{align}
-\ln P(\sigma_0, \alpha, \fknee|\n) = -& \ln P(\sigma_0, \alpha, \fknee)
\notag\\
+& \sum_{\nu>0} \left[
\frac{p_\nu}{N_\nu}+\ln N_\nu
\right].
\end{align}
Here $p_\nu$ are the power spectrum components of the data $\n$, while $N_\nu=N_{\nu}(\sigma_0, \alpha, \fknee)$ is the covariance matrix which in Fourier space is diagonal and given by Equation~\ref{eq:Nnu}. The first term on the right-hand side is a user-defined prior.

To sample from this distribution, we use a standard Metropolis sampler
with a Gaussian proposal density \citep[e.g.,][]{liu:2001}. Each chain
is initialized at the maximum-posterior point, which is found by a
non-linear quasi-Newton search, and the covariance matrix of the
Gaussian proposal density is taken to be the square root of the
curvature matrix, evaluated at the maximum-posterior point. The
elements of the inverse curvature matrix, $\mathcal{C}^{-1} = -
\partial^2 \log P(\theta|\n)/\partial \theta_i \partial \theta_j$, read
\begin{align}
&\mathcal{C}^{-1}_{\theta_i\theta_j} = \notag\\
&\sum_{\nu>0}\left[
\left(\frac{1}{N_\nu}-\frac{p_\nu}{N_\nu^2}\right)
\frac{\partial^2 N_\nu}{\partial \theta_i \partial \theta_j}
+\left(\frac{2p_\nu}{N_\nu^3}-\frac{1}{N_\nu^2}\right)
\frac{\partial N_\nu}{\partial \theta_i} \frac{\partial N_\nu}{\partial \theta_j}
\right].
\end{align}

We have found that this proposal density leads to a Markov chain
correlation length of about 20 samples for typical parameter values,
and we therefore thin our chains by this amount. In addition, we
remove a few post-thinned samples at the beginning of the chain to
remove potential burn-in, although we have never seen evidence of any
such effects. Finally, if we only want to find the maximum-posterior
point, and not run a full-blown Gibbs chain, as discussed in
Section~\ref{sec:posterior}, we terminate the process directly after
the initial quasi-Newton search.

\subsection{Gap filling by constrained realizations}
\label{sec:gap_filling}

\begin{figure}[t]
\mbox{\epsfig{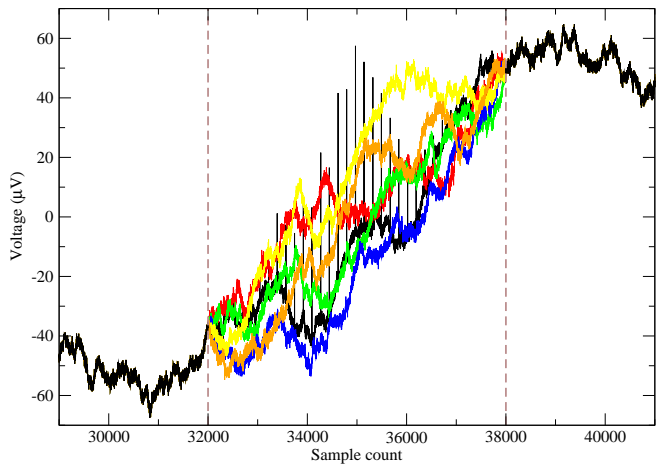}}
\caption{Constrained realizations through a gap in the data. The extent
  of the gap is indicated by the two vertical dashed lines; the solid
  black curve shows the input data, which is dominated by a point source
  (thin spikes) in the masked out region. Five different constrained
  realizations are shown by colored lines. These depend only on the data
  outside the mask, and are therefore not affected by the point source.}
\label{fig:constrained_realizations}
\end{figure}

Next, we need to establish a sampling algorithm for $P(\n,\m|\s, \a,
\theta, \d) = P(\n,\m|\r,\theta)$, where the residual $\r = \d - \P\s
- \T\a = \n + \m$.  Note that $\r$ and $\n$ differ only by $\m$,
which represents contaminated segments of data that are masked out.
The problem is therefore reduced to sampling the components of a sum
of Gaussians given the sum itself. This may be achieved efficiently by
solving the equation 
\begin{align}
\left[\iN+\iM\right]\n &= \iM\r + \N^{-\frac12}\v + \M^{-\frac12}\w \label{eq:gapdraw}
\end{align}
for $\n$, where $\v$ and $\w$ are vectors of standard $N(0,1)$
Gaussian random variates \citep[e.g.,][]{jewell:2004, wandelt:2004,
  eriksen:2004}.

The matrices involved here are typically of the order $10^5 \times
10^5$, so the equation cannot be solved by brute force. But since $\M$
is diagonal in time domain and $\N$ is diagonal in frequency domain
(due to its Toepliz nature), all multiplications are cheap in either time
or Fourier domain, and the equation can be efficiently solved with the Conjugate
Gradients method, properly changing basis as needed.

\begin{figure*}[t]
\mbox{\epsfig{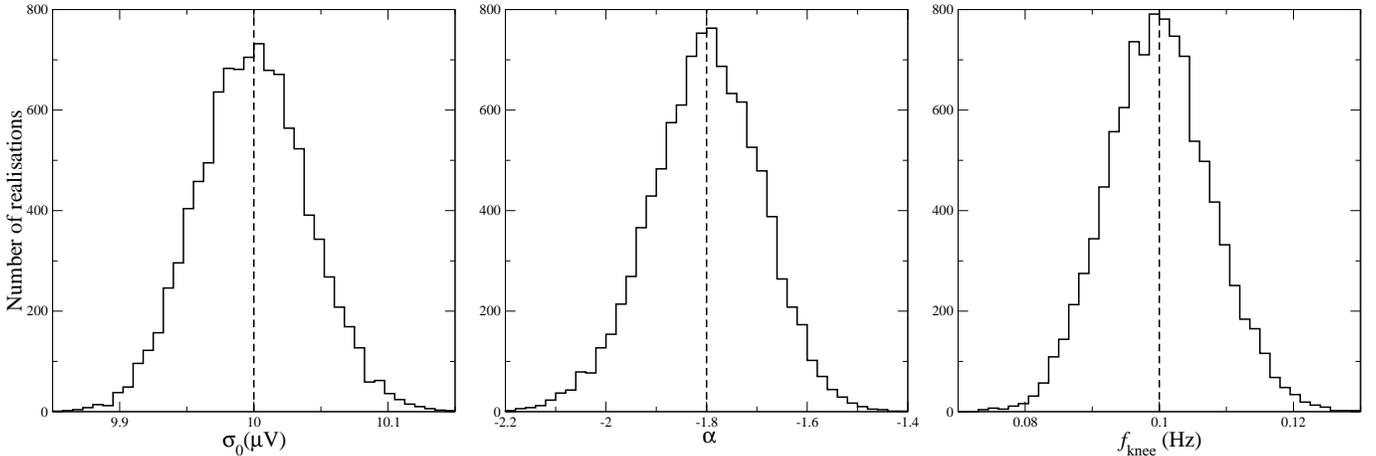}}
\caption{Recovered noise parameters from 10\,000 noise-only
  simulations with gaps. The dashed horizontal lines indicate the true
  input value.}
\label{fig:biastest1}
\end{figure*}

However, there is one practical complication involved in
Equation~\ref{eq:gapdraw}: Since we intend to let $\epsilon\rightarrow
0$ and $a \rightarrow \infty$, the matrix $\iN+\iM$ becomes infinitely
poorly conditioned. We can solve this problem, as well as
significantly simplifying the equation, by splitting it into one
equation for the masked region, and one for the unmasked
region. Introducing the notation $\mathbf{x}_1$ and $\mathbf{x}_2$ for
the unmasked and masked subsets of a vector $\mathbf{x}$ respectively,
we find
\begin{align}
(\iN\n)_1 + \epsilon^{-1}\n_1 &= \epsilon^{-1}\r_1 + (\N^{-\frac12}\v)_1 + \epsilon^{-\frac12}\w_1 \\
(\iN\n)_2 + a^{-1}\n_2 &= a^{-1}\r_2 + (\N^{-\frac12}\v)_2 + a^{-\frac12}\w_2
\end{align}
which in the limit $\epsilon\rightarrow 0$, $a\rightarrow\infty$ simplifies to
\begin{align}
\n_1 &= \r_1 \label{eq:crfinal1}\\
(\iN\n)_2 &= (\N^{-\frac12}\v)_2 \label{eq:crfinal2}
\end{align}
Note that equations~\ref{eq:crfinal1}--\ref{eq:crfinal2}
form an asymmetric equation system\footnote{
Equations~\ref{eq:crfinal1}--\ref{eq:crfinal2}
 can be rewritten as
\[
\left[\begin{array}{cc}
1 & 0 \\ N^{-1}_{21} & N^{-1}_{22}
\end{array}\right] = \left[\begin{array}{c} \r_1 \\ (N^{-\frac12}\v)_2
\end{array}\right]
\]This is asymmetric because we multiplied the unmasked part of the equation
by $\epsilon$ in order to get a finite result.}.  The Conjugate
Gradients method is therefore not directly applicable, and the more
general Biconjugate Gradients method must be used instead.

We found a simple diagonal preconditioner with value $\mathrm{Var}(\n)$ inside
the mask and value 1 outside it to be sufficient.

\subsection{Marginalization over fixed templates}
\label{sec:templates}

The sampling algorithms described in Sections
\ref{sec:noise_estimation} and \ref{sec:gap_filling} defines the core
noise Gibbs sampler, and together form a well-defined and complete
noise estimation algorithm for low signal-to-noise data with
gaps. However, one of the main advantages of the Gibbs sampling
algorithm compared to other alternatives is its natural support for
marginalization over nuisance parameters. In this section we describe
the sampling algorithm for marginalization over fixed time-domain
templates, describing for instance ground pickup, diffuse foregrounds
or cosmic ray glitches.

First, we note that while the original data stream, $\d$, may contain
gaps, and the Toeplitz nature of the noise covariance matrix is in
that case broken, the constrained realization produced in
Section~\ref{sec:gap_filling} restores the Toeplitz symmetry. It is
therefore computationally advantageous to use $\q = \d -\P\s -\m =\n
+\T\a$ as the data in this step\footnote{Of course, one could write
  down a sampling algorithm for $\a$ that only uses the non-masked
  data directly, but this would require heavy time-domain
  convolutions, and not take advantage of the symmetries inherent in
  the noise covariance matrix.}, so that $P(\a,\n|\theta, \m, \s, \d)
= P(\a,\n|\q, \theta)$.

Starting with Equation~\ref{eq:likelihood}, solving for $\a$ and
completing the square in the exponential, the appropriate conditional
distribution for $\a$ is found to be the well-known distribution
\begin{equation}
P(\a|\q, \theta) \propto e^{-\frac{1}{2} (\a-\hat{\a})^t
  (\T^t\N^{-1}\T) (\a-\hat{\a})},
\end{equation}
where $\hat{\a} = (\T^t \N^{-1}\T)^{-1} \T^t \N^{-1}\q$; that is,
$P(\a|\q, \theta)$ is a Gaussian distribution with mean $\hat{\a}$ and
covariance $\C_{\a} = (\T^t\N^{-1}\T)^{-1}$. The same result has been
derived for numerous other applications, one of which was described by
\citet{eriksen:2004}, outlining template amplitude sampling with CMB
sky map data.

Sampling from this distribution is once again straightforward. If one
only wish to marginalize over a small number of templates, the easiest
solution is simply to compute both $\hat{\a}$ and $\C_{\a}$ by
brute-force, and let $\a^{i+1} = \hat{\a} + \L_{\a} \eta$, where
$\L_{\a}$ is the Cholesky factor of $\C_{\a} = \L_{\a}\L_{\a}^t$, and
$\eta$ is a vector of uncorrelated standard Gaussian $N(0,1)$
variates. On the other hand, if there are more than, say, 1000
templates involved, it may be faster to solve the following equation
by Conjugate Gradients,
\begin{equation}
(\T^t \N^{-1} \T) \a = \T^t \N^{-1} \q + \T^t \N^{-\frac{1}{2}} \omega,
\end{equation}
where $\omega$ is a full time-stream of $N(0,1)$ random variates. In
this paper, which has the QUIET experiment as its main application, we
only use the former algorithm; for Planck the latter algorithm may be
useful in order to account for frequent and partially overlapping
cosmic ray glitches efficiently.

\subsection{CMB signal marginalization}
\label{sec:signal}

Most CMB experiments are strongly noise-dominated within relatively
short time periods, and need to integrate over the sky for a long time
in order to produce high-sensitivity maps. For instance, for the
first-season QUIET experiment the mean polarized sensitivity of a
detector was 280$\mu\textrm{K}\sqrt{\textrm{s}}$ \citep{quiet:2011},
while the polarized CMB sky has an RMS of $\sim1\mu\textrm{K}$ on the
relevant angular scales. In such cases, it is an excellent
approximation simply to ignore the CMB contribution when estimating
the noise parameters. However, this approximation does not hold for
all experiments, and one particularly important counterexample is
Planck.

In order to marginalize over the CMB signal we need to be able to
sample from $P(\s,\n|\a, \theta, \m, \d) = P(\s,\n|\u, \theta)$, where
the residual is $\u = \d - \m - \F\a = \P\s + \n$. This is again a
case that involves sampling terms of a Gaussian sum given the sum
itself.  The only difference from Section~\ref{sec:gap_filling} is
that a projection operator is involved in this case,
\begin{align}
(\S^{-1} + \P^T\N^{-1}\P)\s &= \P^T\N^{-1}\u + \S^{-\frac12}\v + \P^T\N^{-\frac12}\w
\end{align}
Here $\v$ and $\w$ are vectors of standard normal samples in pixel and
time domain respectively. Note that $\M$ is not involved this time, so
the expression can be used as it is. As before, multiplications
involving $\N$ are efficient in Fourier space due to the Toeplitz
nature of $\N$, but the CMB signal covariance, $\S$, will in general
be dense, depending on the true CMB power spectrum. This makes this
method prohibitively expensive for general asymmetric scanning
strategies. There are, however, circumstances for which also this
multiplication becomes efficient. One obvious case is that of full sky
coverage, where a change to spherical harmonic basis makes $\S$
diagonal.

\begin{deluxetable*}{lccc}
\centering
\tablecaption{Validation by simulations\label{tab:results}}
\tablecomments{Summary of recovered noise parameters from various
  simulated ensembles. Each column indicates the mean and standard
  deviation of the resulting parameter distribution. The top section
  shows results obtained when simply maximizing the posterior, while
  the bottom section shows the results for a full Gibbs sampling
  analysis. Each run in the top section consists of 10000 simulations,
  while each run in the bottom section consists of 5000
  simulations. All runs have been started with the same seed, to
  ensure directly comparable results.}
\tablecolumns{4}
\tablehead{Simulation  & $\sigma_0$ ($10^{-5}$V) & $\alpha$ & $\fknee$ ($10^{-1}$Hz)}
\startdata
\cutinhead{Posterior maximization (absolute parameter values)}
Gaps only                         & $1.000\pm0.003$ &  $-1.80\pm0.06$ &  $1.00\pm0.05$ \\
Gaps + uncorrected CMB            & $1.000\pm0.003$ &  $-1.80\pm0.06$ &  $1.00\pm0.05$ \\  
Gaps + uncorrected ground pickup  & $1.006\pm0.003$ &  $-1.52\pm0.05$ &  $1.51\pm0.08$ \\
Gaps + corrected ground pickup    & $1.000\pm0.003$ &  $-1.80\pm0.06$ &  $1.00\pm0.05$ \\  
\cutinhead{Gibbs sampling (normalized parameter values)}
Gaps only                         & $-0.02\pm1.00$ &  $0.01\pm1.01$ &  $0.03\pm1.01$ \\
Gaps + uncorrected CMB            & $-0.02\pm1.00$ &  $0.00\pm1.01$ &  $0.02\pm1.01$ \\
Gaps + uncorrected ground pickup  & $\phantom{-}1.78\pm1.03$ &  $6.04\pm1.36$ &  $6.63\pm0.77$ \\
Gaps + corrected ground pickup    & $-0.02\pm1.00$ &  $0.00\pm1.00$ &  $0.03\pm1.00$ 
\enddata
\end{deluxetable*}

More interestingly, $\S$ also becomes diagonal when expressed in
Fourier basis on a circle on the sky. To see this, consider a circle
with radius $\theta$ on the sphere, parametrized by the angle $\phi$.
The covariance between the points $p_1 = (\theta,\phi_1)$ and $p_2 =
(\theta,\phi_2)$ with angular distance $r$ is given by the two-point
correlation function:
\begin{align}
S(p_1,p_1) = C(r) = \frac{1}{4\pi} \sum_{l=0}^\infty (2l+1)C_l P_l(cos(r))
\end{align}
Since $r(p_1,p_2)$ for the case of a circle only depends on
$\Delta\phi$, and is independent of $\phi$ itself, $S(\phi_1,\phi_2)$
is a Toepliz matrix, and is therefore diagonal in Fourier space. This
is highly relevant for Planck, since the Planck scanning strategy
\citep{planck:2011} naturally divides into scans of
circles. Therefore, in this case sampling $\s \leftarrow P(\s|\u)$ can
be done at very low extra cost.

\section{Application to simulated data}
\label{sec:quiet}

In this section we demonstrate the noise Gibbs sampler as described in
Section~\ref{sec:algorithm} on a particular type of QUIET
simulations. QUIET is a radiometer-based CMB B-mode polarization
experiment located in the Atacama desert \citep{quiet:2011}, which
took observations from August 2008 to December 2010. The first results
were based on only nine months worth of data, and yet already provided
the second most stringent upper limit on the tensor-to-scalar ratio,
$r$, to date based on CMB polarization measurements.

In its normal mode of operation, QUIET observed four separate CMB
fields on the sky which were each chosen because of their low
foreground levels. In this mode, the experiment is totally noise
dominated on time scales of an hour or less, with a mean polarized
sensitivity per detector diode of 280$\mu\textrm{K}\sqrt{\textrm{s}}$
\citep{quiet:2011}.

However, QUIET also observed two Galactic patches, one of which was
the Galactic center, as well as several bright calibration objects,
such as the Moon, Tau Alpha and RCW38. These sources are bright enough
to be seen visually in each detector time stream, and they can
therefore bias any noise estimates unless properly accounted for. Such
objects also complicate automated data selection processes, since it
is difficult to distinguish between an astrophysical object and an
instrumental glitch.

In this section we show how the algorithm developed in
Section~\ref{sec:algorithm} may be applied to such
situations. Specifically, we consider a observing session lasting for
about 40 minutes of a field including a bright source with known
location, and assume that the data may be modeled as $\d = \n + \T\a
+ \m$. Here $\T$ is a single template describing possible sidelobe
pick-up from the ground, constructed from the full observing season as
described by \citet{quiet:2011}, and $\m$ is a time-domain mask that
removes any samples that happen to fall closer than $1^{\circ}$ from
the source center. The total number of samples in the data stream is
60\,949, and the total number of masked samples is 651.


The simulations used in this section are constructed as follows. We
set up an ensemble of $10^4$ time streams containing correlated
Gaussian random noise with $\sigma_0 = 10^{-5}\textrm{V}$,
$\alpha=-1.8$ and $f_{\textrm{k}} = 0.1\textrm{Hz}$; the white noise
and spectral index are representative for a QUIET detector, while the
knee frequency is grossly exaggerated to push the algorithm into a
difficult region of parameter space, as well as to more clearly
visualize the outputs of the algorithm. A far more reasonable value
for QUIET is $f_{\textrm{k}} = 10\textrm{mHz}$, and we have of course
verified that the algorithm also works for such cases. Further, it
reaches convergence faster in that case than with the extreme value of
$f_{\textrm{k}}$ used in the present simulations.

\subsection{Visual inspection of constrained realizations}

Before considering the statistical properties of the resulting
posterior distributions, it is useful to look visually at a few
constrained realizations in order to build up intuition about the
algorithm. In order to highlight the behavior of the constrained
realizations, we make two adjustments to the above simulation
procedure for this case alone: First, we replace the tuned mask with a
wide 6000-sample mask, covering the entire time range in which the
source is visible, and second, we make the correlated noise component
stronger by setting $\sigma_0=10^{-6}\textrm{V}$, $\fknee = 1\textrm{Hz}$ and
$\alpha=-2.3$.

The results are shown in
Figure~\ref{fig:constrained_realizations}. The raw data are shown in
the solid black line, and the vertical dashed lines indicate the
extent of the gap. The colored curves within the gap shows 5
difference constrained realizations; note that together with the black
solid curve outside the mask, any of these form a valid noise
realization with the appropriate noise power spectrum as defined by
$\sigma_0$, $\alpha$ and $\fknee$. They are each a valid sample drawn
from $P(\n, \m|\d)$. However, if one had tried to estimate the noise
spectrum also using the data inside the gap, the source signal (seen
as sharp spikes in Figure~\ref{fig:constrained_realizations}) would
bias the resulting noise parameters.

In this paper, we consider the constrained realizations primarily to
be a useful tool that allows for fast noise covariance matrix
multiplications in Fourier space. However, these constrained
realizations can of course also be useful in their own right, for
instance for deglitching a time stream before map making. 

\subsection{Validation and statistical characterization}

\begin{figure*}[t]
\mbox{\epsfig{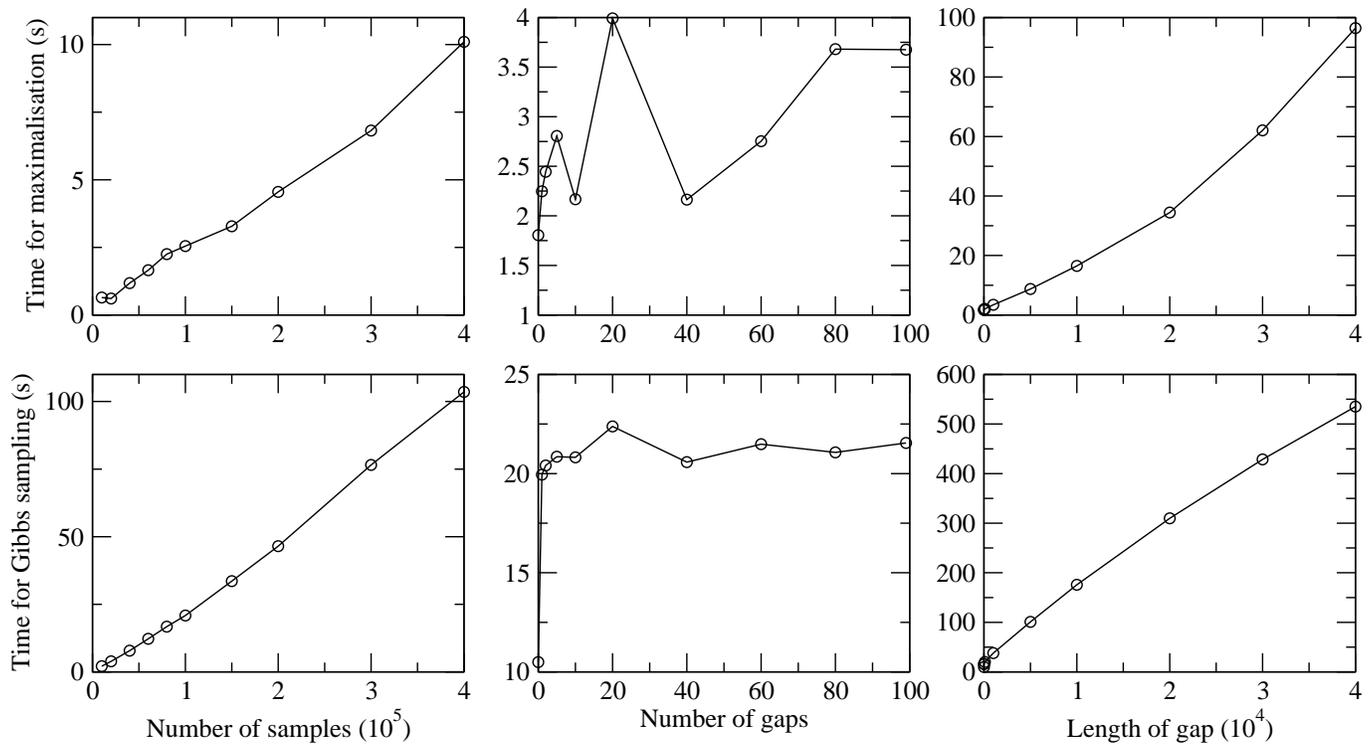}}
\caption{\emph{Left:} CPU time as a function of total number of
  samples in the time stream, keeping the number of masked samples
  constant. \emph{Middle:} CPU time as a function of number of masked
  samples. The size of each gap is kept constant at 100 samples, and
  only the number of gaps vary. \emph{Right:} CPU time as a function
  of number of masked samples, but in this case there is only one gap
  of varying width.}
\label{fig:CPUtime}
\end{figure*}

We now seek to statistically validate our algorithms and codes. Both
the posterior maximization and the Gibbs sampling algorithms are
considered. The number of simulations are 10\,000 for posterior
maximization and 5000 for Gibbs sampling, with properties as described
above. In each case, we consider four different models. First, we
analyze simulations including only noise and gaps. Second, we add a
CMB signal to each realization, but do not attempt to correct for
it. Third, we add a strong ground template to each realization, and do
also not attempt to correct for it. Fourth, we analyze the same
ground-contaminated simulations as above, but this time do marginalize
over an appropriate template. The same random seeds were used in each
of the four simulation and analysis results, in order to allow for
direct comparison of results between runs.
 
The results from this exercise are summarized in
Table~\ref{tab:results}, and histograms for the first case are shown
in Figure~\ref{fig:biastest1}. First, the the upper section in
Table~\ref{tab:results} lists the mean and standard deviation of the
recovered parameters for the posterior maximization algorithm. Recall
that the input parameters were $(\sigma_0, \alpha, \fknee) =
(10^{-5}\textrm{V}, -1.8, 0.1\textrm{Hz})$, and these are recovered
perfectly in all cases, except for the one involving an uncorrected
ground template, as expected.

Second, in the bottom half we show the results from the Gibbs sampling
analyses, but this time in terms of normalized parameters on the form
$r = (\theta_{\textrm{est}}-\theta_{\textrm{in}}) /
\sigma_{\textrm{est}}$, where $\theta_{\textrm{est}}$ and
$\sigma_{\textrm{est}}$ are the mean and standard deviation of the
Gibbs chain for a given parameter (removing the first 10\% of the
samples for burn-in), and $\sigma_{\textrm{in}}$ is the true input
value. If the Gibbs chain is both unbiased and has the correct
dispersion, $r$ should be Gaussian distributed with zero mean and unit
variance. As seen in Table~\ref{tab:results}, this is indeed the case.

We also note that adding a CMB component to these simulations do not
bias the noise estimates, simply because the CMB is too weak to be
detected on the time scales considered here. This confirms the
assumption made by the QUIET team when estimating their noise
properties: The QUIET observations are sufficiently noise dominated on
a one-hour time scale that the CMB can be safely neglected for noise
estimation purposes.

\subsection{Resource requirements}

To be practical, it is not sufficient that a method is robust and
accurate, but it must also be computationally efficient. For the
present algorithm the two most important parameters for computational
speed are 1) the total number of samples in the time stream, $n$, and
2) the number of masked samples, $m$, while also the relative position
of the masked samples play an important role.

In Figure~\ref{fig:CPUtime} we show the scaling of each of the two
algorithms (posterior maximization and Gibbs sampling) as a function
of both $n$ (left panel) and $m$ (right panel). In the left plot,
$m$ was fixed at 1000, divided into ten gaps of 100 samples each, and
only the total length of the data stream was varied. In this case, we
should expect the scaling of the overall algorithm to be dominated by
Fourier transforms, suggesting an overall behavior given by
$\mathcal{O}(n \log n)$. As seen in Figure~\ref{fig:CPUtime}, this
approximation holds to a very high degree, both for posterior
maximization and Gibbs sampling. Further, we see that the CPU time
required to analyze a single 100\,000 sample data set with 1000
samples removed is 3 seconds for posterior maximization and 21 seconds
for Gibbs sampling. 

In the middle panel, we fix $n$ at 100\,000, and increase $m$ by
varying the number of gaps, each extending 100 samples. Perhaps
somewhat surprisingly, we see that the computing time in this case is
nearly independent of $m$. The reason for this is simply that the
number of conjugate gradient iterations required for the gap filling
procedure is largely determined by condition number (ie., the ratio
between the highest and smallest eigenvalue) of the covariance matrix
of a single gap. Having more gaps separated by more than one
time-domain correlation length effectively corresponds to performing
multiple matrix inversions in parallel, and the net cost therefore do
not increase significantly.

In the third panel, we increase $m$ by making one gap larger, as
opposed to adding many small gaps. In this case the CPU time does
increase dramatically, because it becomes increasingly hard for the
algorithm to fill the missing pieces of the data stream. In this case,
the noise covariance matrix condition becomes larger and larger. 

\section{Summary}
\label{sec:conclusions}

We have described and implemented a Bayesian framework for estimating
the time-domain noise power spectrum for non-ideal CMB
experiments. This framework is conceptually identical to a previously
described method for estimating the angular CMB power spectrum from
CMB sky maps \citep{jewell:2004, wandelt:2004, eriksen:2004}, and
relies heavily on the Gibbs sampling algorithm. The single most
important advantage of this method over existing competitors in the
literature derives from the conditional nature of the Gibbs sampler:
Additional parameters may be introduced \emph{conditionally} into the
algorithm. This allows for seamless marginalization over nuisance
parameters, which otherwise may be difficult to integrate. A second
important advantage of the method is the fact that it provides proper
uncertainties on all estimated quantities, which at least in principle
later may be propagated into final estimates of the uncertainties of
the CMB sky map and angular power spectra.

In this paper we implemented support for two general features that are
useful for analysis of realistic data, namely constrained realizations
and template sampling. The former is useful whenever there are gaps in
the data, for instance due to an instrumental glitch, or there are
strong localized sources in the sky that may bias the noise estimate:
In these cases, the gaps are refilled with a constrained noise
realization with the appropriate noise parameters, such that the full
time stream represents a proper sample from a Gaussian distribution
with a noise covariance matrix, $\N$. Since the time stream no longer
contains gaps, the Toeplitz symmetry of the noise covariance matrix is
restored, and matrix multiplications may be performed quickly in
Fourier space.

The second operation, template sampling, is also a powerful and
versatile technique for mitigating systematic errors. In this paper we
mostly focused on data from the QUIET experiment, for which ground
pickup from sidelobes is one significant source of systematics
\citep{quiet:2011}. In a future publication we will apply the same
method to simulations of the Planck experiment, for which cosmic ray
glitches is an important source of systematic errors. As detailed by
\citet{planck_hfi:2011}, these cosmic rays may be modeled in terms of
a limited set of time domain templates, and the algorithms presented
in this paper should therefore prove useful for mitigating the effects
of these glitches, as well as for propagating the corresponding
uncertainties into the final noise spectrum parameters.

\begin{acknowledgements}
We thank the QUIET collaboration for stimulating discussions. The
computations presented in this paper were carried out on Titan, a
cluster owned and maintained by the University of Oslo and NOTUR. This
project was supported by the ERC Starting Grant StG2010-257080. Some
of the results in this paper have been derived using the HEALPix
\citep{gorski:2005} software and analysis package.
\end{acknowledgements}

\end{document}